\documentclass[preprint]{aastex631}
\usepackage{times}
\usepackage{ulem}
\usepackage{bm}  

\newcommand{\sdo}{{\it SDO}} %

\newcommand{\Boyang}[1]{\textcolor{blue}{#1}}

\shorttitle{Characterizing High-latitude Modes}
\shortauthors{Ding et al.}

\begin{document}
\title{Characterizing the Observational Properties of the Sun's High-latitude $m=1$ Inertial Mode}
\author[0009-0001-5405-4042]{Boyang Ding}
\affiliation{Department of Physics, Stanford University, Stanford, CA 94305-4013}
\affiliation{W.~W.~Hansen Experimental Physics Laboratory, Stanford University, Stanford, CA 94305-4085, USA}

\author[0000-0002-6308-872X]{Junwei Zhao}
\affiliation{W.~W.~Hansen Experimental Physics Laboratory, Stanford University, Stanford, CA 94305-4085, USA}

\author[0000-0002-2632-130X]{Ruizhu Chen}
\affiliation{W.~W.~Hansen Experimental Physics Laboratory, Stanford University, Stanford, CA 94305-4085, USA}

\author[0000-0003-2678-626X]{Matthias Waidele}
\affiliation{W.~W.~Hansen Experimental Physics Laboratory, Stanford University, Stanford, CA 94305-4085, USA}

\author[0000-0003-1753-8002]{Sushant S. Mahajan}
\affiliation{W.~W.~Hansen Experimental Physics Laboratory, Stanford University, Stanford, CA 94305-4085, USA}

\author[0000-0001-6754-1520]{Oana Vesa}
\affiliation{W.~W.~Hansen Experimental Physics Laboratory, Stanford University, Stanford, CA 94305-4085, USA}

\begin{abstract}
Low-$m$ inertial modes have been recently discovered in the Sun’s high-latitude regions. 
In this study, we characterize the observational properties of the $m=1$ mode by analyzing time-distance subsurface flow maps. 
Synoptic flow maps, constructed from daily subsurface flow maps using a tracking rate corresponding to the rotation at latitude $65\degr$, are filtered in both the spherical harmonic and Fourier domains to retain only the $m=1$ mode and its dominant frequencies. 
Our analysis reveals a power distribution that is significantly stronger in the northern polar region.
The mode’s power exhibits an anti-correlation with solar activity, remaining strong and persistent during the solar activity minimum and becoming weaker and more fragmented during the solar maximum. 
Magnetic flux transported from low to high latitudes influences both the mode’s power and lifetime, enhancing its power and shortening its lifetime upon arrival. 
The phases of the $m=1$ mode in the northern and southern polar regions are near-antisymmetric for most of the time with short deviations. 
We also compute zonal and meridional phase velocities of the mode and find that it exhibits significantly less differential rotation than its surrounding plasma.
The meridional phase velocity, comprising both the local plasma’s meridional flow and the mode's intrinsic phase motion, is directed poleward below latitude $70\degr$ and equatorward above this latitude. 
These observational findings underscore the need for a deeper understanding of the internal dynamics of the low-$m$ modes, which may offer valuable insights into the structure and dynamics of the solar interior.
\end{abstract}


\section{Introduction}
\label{sec1}

Recent discoveries of Rossby waves and high-latitude inertial modes on the Sun have reignited interest in these phenomena, both observationally and theoretically. 
\citet{Loptien18} reported the detection of equatorial Rossby waves in a latitudinal band within $\pm20\degr$ from the equator by analyzing the power spectrum of large-scale radial vorticities derived from surface horizontal flow fields. 
These sectoral modes, restored by the Coriolis force, were found to approximately follow the dispersion relation, $\omega = - 2 \Omega / (m+1)$ (where $\omega$ is the angular frequency of the Rossby waves, $\Omega$ is the rotation rate of the Sun, and $m$ is the azimuthal order).
Subsequent observations using subsurface flow fields obtained from ring-diagram helioseismology \citep{Proxauf20, Hanson20}, time-distance helioseismology \citep{Liang19, Waidele23}, and other helioseismic techniques, such as mode coupling \citep{Hanasoge19, Mandal21}, further confirmed the detection of these waves and expanded our understanding of the phenomenon. 
Notably, the power of equatorial Rossby waves has been observed to show a positive correlation with sunspot numbers, becoming stronger during solar maximum \citep{Waidele23, Lekshmi24}.

Among the family of solar inertial modes \citep{Gizon21}, the high-latitude low-$m$ modes (specifically, $m=1$ or $2$) are the easiest to detect due to their relatively large longitudinal and latitudinal velocities. 
These modes were first reported by \citet{Hathaway13} using supergranular tracking and were initially identified as giant convective cells \citep{Hathaway21}. 
Using flow fields from various helioseismic techniques, \citet{Bogart15} and \citet{Zhao16} confirmed these long-living and large-scale structures. 
However, by comparing observations with the normal modes of a rotating spherical shell, \citet{Gizon21} reclassified these high-latitude structures as low-$m$ inertial modes.
Subsequent analyses using longer observational periods and multiple data sources revealed that the power of the $m=1$ mode also varies with the solar cycle but with an anti-correlation, becoming weaker during solar maximum \citep{Gizon24, Liang24}, in contrast to the equatorial Rossby waves.

As observational evidence continues to accumulate, theoretical and numerical approaches advance rapidly to enhance our understanding of the various types of inertial modes. 
Numerous studies have focused on inertial modes or thermal Rossby waves through numerical simulations or solving for eigenfunctions \citep[e.g.,][]{ Hindman22, Triana22, Jain23, Philidet23, Blume24}.
Here, we narrow our focus to the few numerical studies that are most relevant to the study presented in this paper.
By solving eigenfunctions in a two-dimensional differentially rotating convection zone, \citet{Gizon21} and \citet{Bekki22} identified numerous inertial modes, including equatorial Rossby waves and high-latitude low-$m$ inertial modes. 
Their work revealed that these modes are highly sensitive to turbulent viscous diffusivity and the superadiabaticity of the convection zone, suggesting that observations of inertial modes could provide crucial constraints on these internal parameters. 
\citet{Bekki24} further proposed that high-latitude baroclinically unstable inertial modes may play a key role in shaping the Sun’s differential rotation.
More recently, \citet{Dikpati24} numerically investigated the interaction between the Sun’s surface magnetic fields and their poleward migration with high-latitude low-$m$ inertial modes. 
Their results suggest that magnetic migration may positively influence the formation of these modes.

In this paper, we analyze high-latitude low-$m$ inertial modes using subsurface flow fields derived from the {\it Solar Dynamics Observatory} / Heliosesimic and Magnetic Imager \citep[\sdo/HMI;][]{Pesnell12, Scherrer12, Schou12} time-distance helioseismic data-analysis pipeline \citep{Zhao12}. 
We characterize the $m=1$ mode by analyzing their frequency- and time-dependent power distributions in both polar regions; exploring their relationship with local magnetic fields, monthly sunspot numbers, and zonal flow; measuring their phase speeds in both the zonal and meridional directions; and examining the phase relations between the modes in two polar regions.
This paper is structured as follows: we introduce our data preparation in Section~\ref{sec2}, present our analysis method and results in Section~\ref{sec3}, and discuss the results and give conclusions in Section~\ref{sec4}.

\section{Data Preparation}
\label{sec2}

Longitudinal velocities $v_\phi (\theta, \phi)$, where $\theta$ represents latitude and $\phi$ longitude, are defined as velocities parallel to the direction of solar rotation.
The subsurface $v_\phi (\theta, \phi)$ obtained from the \sdo/HMI time-distance helioseismic analysis pipeline \citep{Zhao12} for the depth of $0-1$\,Mm are used in this study. 
These flows, with a spatial sampling rate of $0\fdg12$\,pixel$^{-1}$, are generated every 8 hr using \sdo/HMI Dopplergrams and span approximately $\pm80\fdg4$ latitude near the central meridian. 
Such flow maps have been continuously available since May 2010, lasting already  15\,yr, and are ideal for long-term studies of the inertial modes. 
In this study, we use the data from 2010 May 19 to 2024 August 23, spanning a total of 14 years and 3 months and covering Carrington Rotation (CR) 2097 -- CR2288.
These subsurface flows from time-distanced helioseismology pipeline typically have a higher temporal cadence and finer spatial sampling than those derived from the ring-diagram pipeline \citep{Bogart11} and used by, e.g., \citet{Gizon21}. 
However, since inertial modes have very large spatial scales and evolve slowly over time, these different datasets are expected to yield similar results.

\begin{figure}[!t]
\includegraphics[width=1.0\textwidth]{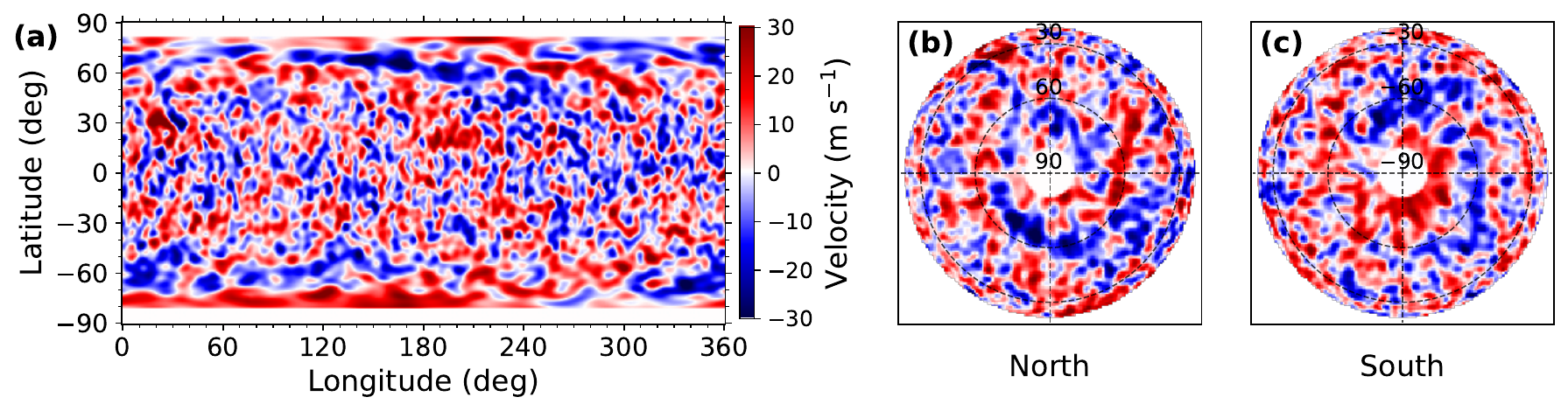}
\caption{(a) An example of synoptic longitudinal-flow maps, $v_\phi (\theta,\phi)$, after applying a progressive Gaussian smoothing on the data and removing the mean zonal flow. 
(b) Same synoptic flow map viewed from above the north pole, constructed using the orthographic projection.
(c) Same as (b) but for the south pole. 
White regions in all panels indicate areas with no data coverage.
\label{fig:data}}
\end{figure}

For this analysis, we do not use the Carrington synoptic subsurface flow maps that result directly from the time-distance helioseismology pipeline. 
Because the Carrington rotation rate is substantially faster than the rotation rate of the high-latitude regions, the Carrington synoptic maps do not provide a complete $360\degr$ longitude coverage of flows at higher latitudes, but cover a substantially smaller portion. 
This makes the spherical harmonic decomposition inaccurate in the high-latitude regions, particularly in the determination of $m$.
To cover one full rotation in the region, as well as to keep the dominant frequencies of low-$m$ modes not far off zero frequency, we track the daily subsurface flow maps and construct new synoptic flow maps using the rotation rate at latitude $65\degr$, which is 365.4\,nHz following \citet{Komm1993}. 
After the construction, we have a total of 151 synoptic flow maps covering the same aforementioned time period. 

Throughout the period, to remove the differential rotation in the $v_\phi$ maps, the $v_\phi$ is averaged as a function of latitude within a moving window that includes five synoptic maps and is subtracted from the middle of the five maps.
In this process, two synoptic maps at the beginning and two at the end of the sequence are dropped, leaving 147 maps for further analysis.
The data are then Gaussian-smoothed to eliminate small-scale structures.
Since the spatial resolution gradually decreases with latitude, a progressive Gaussian smoothing is applied, with the longitudinal size of the two-dimensional Gaussian kernels increasing accordingly with latitude while their latitudinal size kept constant.
The latitudinal sizes, $\sigma_y$, characterized by the standard deviations of the Gaussian kernels, vary with the corresponding longitudinal sizes, $\sigma_x$, following
$\sigma_x = \sigma_y / \cos \theta$. 
Figure~\ref{fig:data} shows an example of the synoptic flow maps constructed using the local rotation tracking rate, as well as two flow maps constructed using orthographic projection\footnote{\url{https://en.wikipedia.org/wiki/Orthographic_projection}} for both the northern and southern polar regions.

\citet{Gizon21} and \citet{Liang24} utilized $v_\phi^+$ and $v_\phi^-$, the symmetrized and antisymmetrized longitudinal velocities of the northern and southern hemispheres.
In this study, we analyze $v_\phi^\mathrm{N}$ and $v_\phi^\mathrm{S}$, the longitudinal velocities in the northern and southern hemispheres, respectively, to emphasize the differences in the wave characteristics between the two hemispheres.
Technically, the latitudinal velocities, $v_\theta (\theta, \phi)$, can also be derived from the same time-distance helioseismology pipeline, but due to the strong center-to-limb effect \citep{Zhao12b}, these velocities become increasingly unreliable beyond approximately latitude $70\degr$ and are therefore not used in this study. 

\section{Data Analysis and Results}
\label{sec3} 

\subsection{Mode Filtering and Power Distribution} 
\label{sec31}

\begin{figure}[!t]
\includegraphics[width=1.0\textwidth]{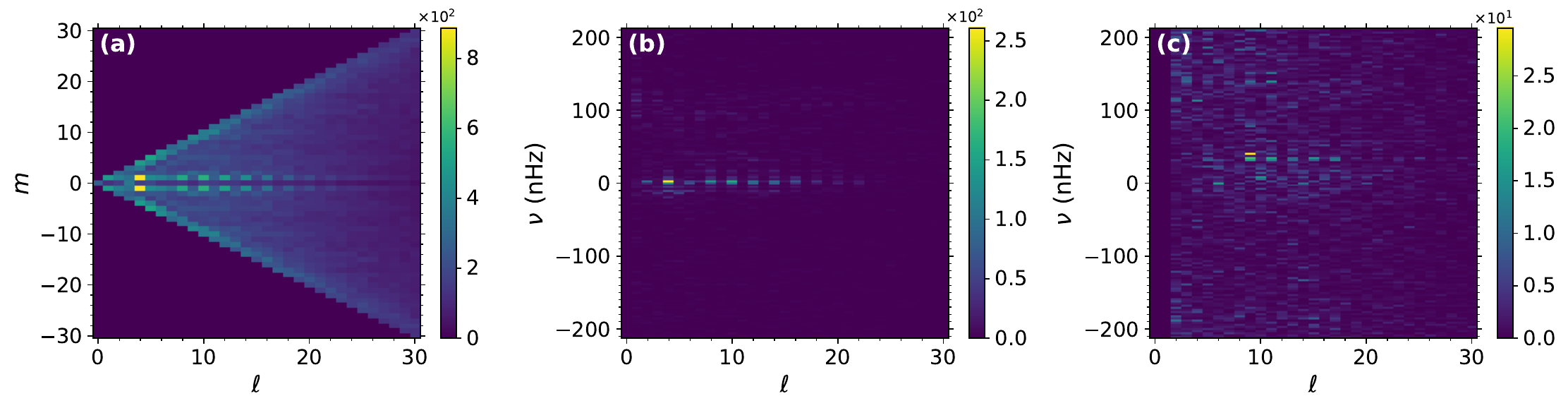}
\caption{(a) Power distribution, summed over all $\nu$'s and displayed as a function of  spherical harmonic parameters $(\ell, m)$. 
(b) Power distribution for $m=1$, displayed with $(\ell,\nu)$.
(c) Power distribution for $m=2$, displayed with $(\ell,\nu)$.
\label{fig:sph_fft}}
\end{figure}

After the $v_\phi(\theta, \phi)$ data are tracked and the progressive Gaussian smoothing is applied, spherical harmonic decomposition is performed separately on each synoptic flow map, yielding $A(\ell,m,t)$, where $\ell$ represents harmonics degree, $m$ the azimuthal order, and $t$ the time. 
For each pair of $(\ell,m)$, Fourier transform is applied on the time sequence of $A(\ell,m,t)$ to convert time $t$ into frequency $\nu$, resulting in $A(\ell,m,\nu)$. 
Figure~\ref{fig:sph_fft}a presents the total power of the inertial modes as a function of $\ell$ and $m$, summed over all $\nu$'s. 
Notably, summing over all $t$'s before performing the Fourier transform yields the same result, consistent with Parseval’s theorem. 
The power is concentrated at $m=1$ and some even-numbered $\ell$'s below $\ell$ of 20. 
The power corresponding to $m=2$ is relatively weak but non-negligible, while for $m \ge 3$, the power is mostly negligible.
Figure~\ref{fig:sph_fft}b and \ref{fig:sph_fft}c show the $\ell - \nu$ power distributions of $m=1$ and $m=2$ modes, respectively. 

\begin{figure}[!t]
\plotone{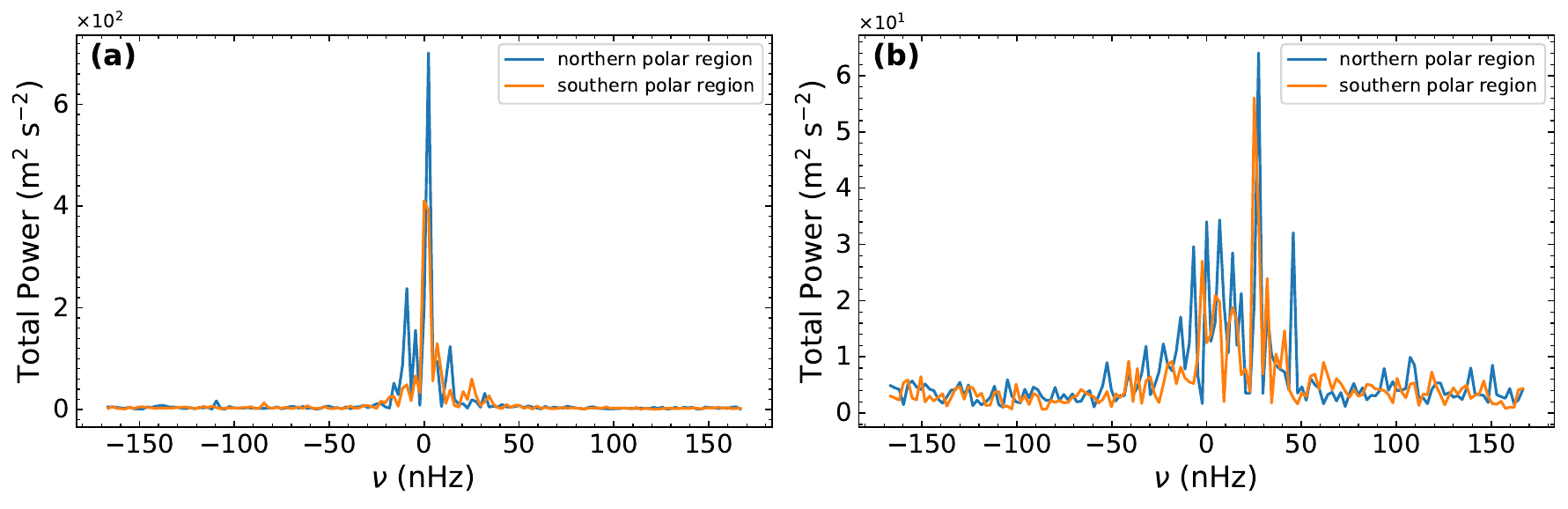}
\caption{(a) Total power of $m=1$ mode, summed over $\ell \le 20$, for both the northern and southern polar regions, displayed as a function of $\nu$.
(b) Total power of $m=2$ mode for both polar regions. \label{fig:spectrum}}
\end{figure}

By summing over all $\ell \le 20$ for $m=1$ in the northern and southern polar regions (in this study, we consider areas between latitudes of $50\degr - 80\degr$ as polar regions) separately, we obtain the power distribution of $m=1$ mode as a function of $\nu$.
As shown in Figure~\ref{fig:spectrum}a, the mode power concentrates near $\nu = 0 $\,nHz, which is relative to the tracking rate, with a width of approximately $\pm20$\,nHz (the frequency resolution is $\sim2.9$\,nHz), consistent with past observations \citep{Gizon21}. 
Notably, the peak power in the northern hemisphere is significantly stronger than in the southern hemisphere.
Additionally, the power in the northern polar region exhibits a secondary peak around $-12$\,nHz, and the total power is stronger than the total power of the southern polar region. 

The power for $m=2$ is also calculated similarly. 
As shown in Figure~\ref{fig:spectrum}b, the peak power occurs at approximately 35\,nHz (32\,nHz) in the northern (southern) polar region relative to the reference frame, with two regions showing similar peak power and frequency span. 

\begin{figure}[!t]
\plotone{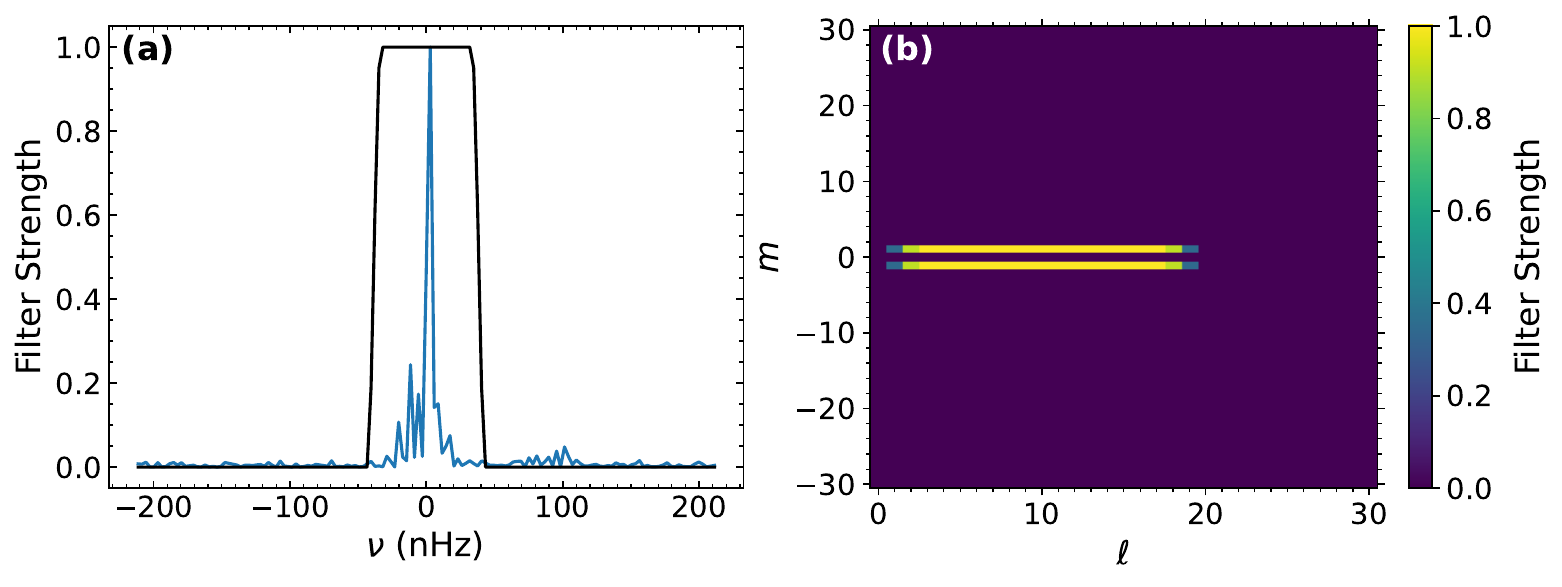}
\caption{(a) Filter (black) applied in the frequency domain, with the normalized power distribution of the ($\ell = 4, m=1$) mode (blue line) in the background as a reference.
The filter has a cosine-bell shape on both ends to reduce the aliasing effect.
(b) Filter applied on $(\ell, m)$, keeping only the power of $m=1$ and $\ell = 0 - 20$, with a cosine-bell on both ends of $\ell$.
\label{fig:filter}}
\end{figure}

This study focuses primarily on the $m=1$ mode and leaves $m=2$ mode for future studies. Filters are applied in the spherical harmonic and the Fourier domains to preserve only the modes of interest. 
As shown in Figure~\ref{fig:filter}\Boyang{a}, a frequency filter, 90\,nHz wide with a cosine bell on either side, is applied on $\nu$.
For the filter in the spherical harmonic domain (shown in Figure~\ref{fig:filter}b), the $m=1$ and $0 \le \ell \le 20$ signals are kept, with cosine bells applied on both ends of the $\ell$ range. 
The combination of the frequency and spherical-harmonic filters will keep only $m=1$ mode left for further analysis.
Once the filters are applied on $\ell$, $m$, and $\nu$, the data are first inverse Fourier transformed into the time domain for each pair of $(\ell,m)$.
Then, for each time step, the inverse spherical harmonic transform is performed to convert the data into the $(\theta,\phi)$ domain. 

\begin{figure}[!t]
\includegraphics[width=1.0\textwidth]{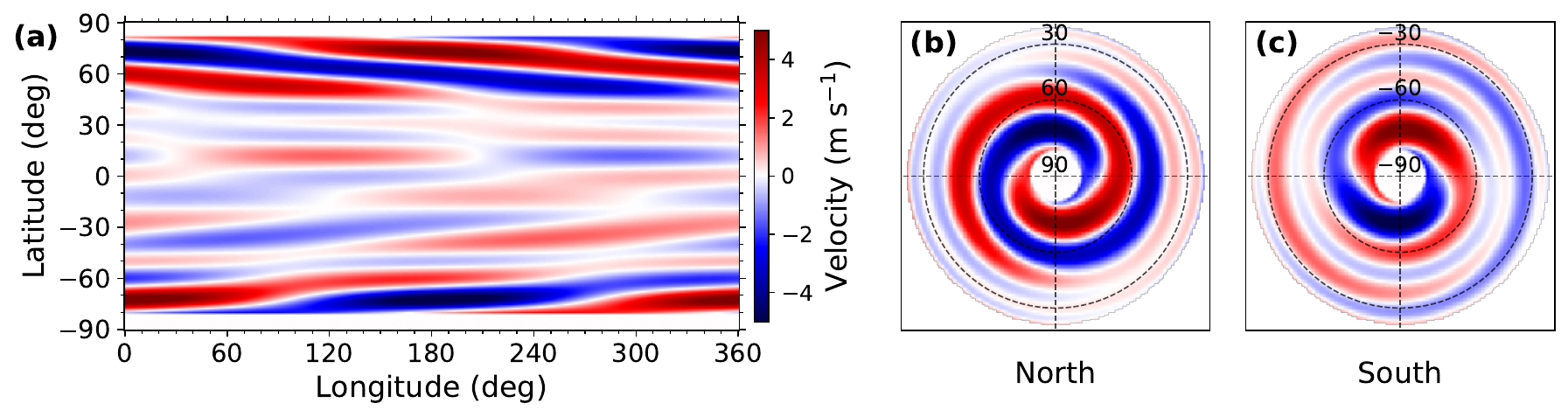}
\caption{(a) An example synoptic longitudinal-flow map after the filter is applied to keep only $m=1$ mode near its dominant frequencies. 
(b) Same synoptic flow map but projected around the north pole.
(c) Same synoptic flow map but projected around the south pole.
An animation accompanying this figure is available online. 
The animated figure shows all 147 flow maps projected around both the North and South poles, displayed consecutively to cover the full analysis period. It dynamically illustrates the evolution of the characteristic $m=1$ yin-yang pattern, with the static images in panels b and c corresponding to a single representative frame from the animation.}
\label{fig:filtered_data}
\end{figure}

Figure~\ref{fig:filtered_data} presents a sample synoptic flow map of the $m=1$ mode, shown in both longitude -- latitude coordinate (panel a) and polar views (panels b and c). 
An animation accompanying Figure~\ref{fig:filtered_data} illustrates the temporal evolution of the $m=1$ mode in both polar regions, and our subsequent analysis largely depends on the filtered data displayed in this animation to further characterize the mode's properties. 
The animation reveals that the yin-yang structure of the $m=1$ mode remains largely stationary for one or two years before transitioning into a new yin-yang pattern, usually accompanied by a significant shift in longitude that is referred to, hereafter, as a phase jump. 
These phase jumps occur when one mode decays and a new mode begins to form at a different longitude, although this transition is sometimes only vaguely visible. 
The lifetime of the mode, from its formation to its decay, ranges from approximately one year to over two years (see Figures~\ref{fig:n_power_t}a and \ref{fig:s_power_t}a), consistent with the mean lifetime reported by \citet{Liang24}.

\subsection{Comparing Modes with Magnetic Field, Sunspot Number, and Zonal Flow}
\label{sec32}

\begin{figure}[!t]
\plotone{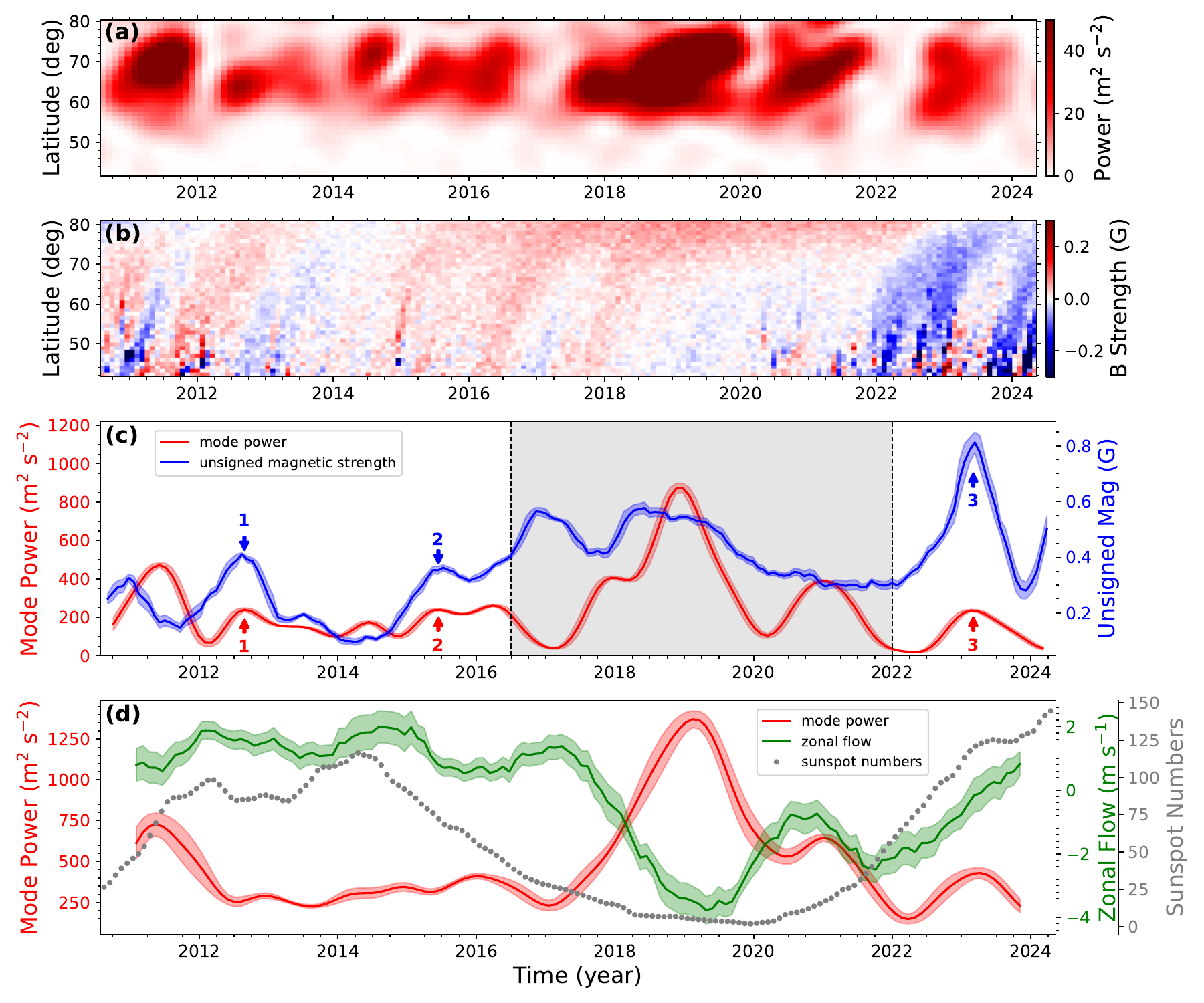}
\caption{Comparisons of mode power with magnetic field, zonal flow, and sunspot number for the northern hemisphere. 
(a) Power of $m=1$ mode, integrated across all longitudes of each rotation for each latitude and displayed as a function of time and latitude.
(b) Magnetic butterfly diagram during the interested time period and covering only the high-latitude regions in the northern hemisphere.
(c) Comparison of integrated mode power and magnetic flux, both smoothed with a 5-rotation running window.
Arrows associated with a number indicate the peaks in mode power and magnetic flux that correspond to each other in the timing of occurrence. 
The shaded region indicates the activity minimum period.
(d) Comparison of integrated mode power, mean zonal flow velocity, and monthly sunspot number, all of which are smoothed with a 13-month running window.
\label{fig:n_power_t}}
\end{figure}

\begin{figure}[!t]
\plotone{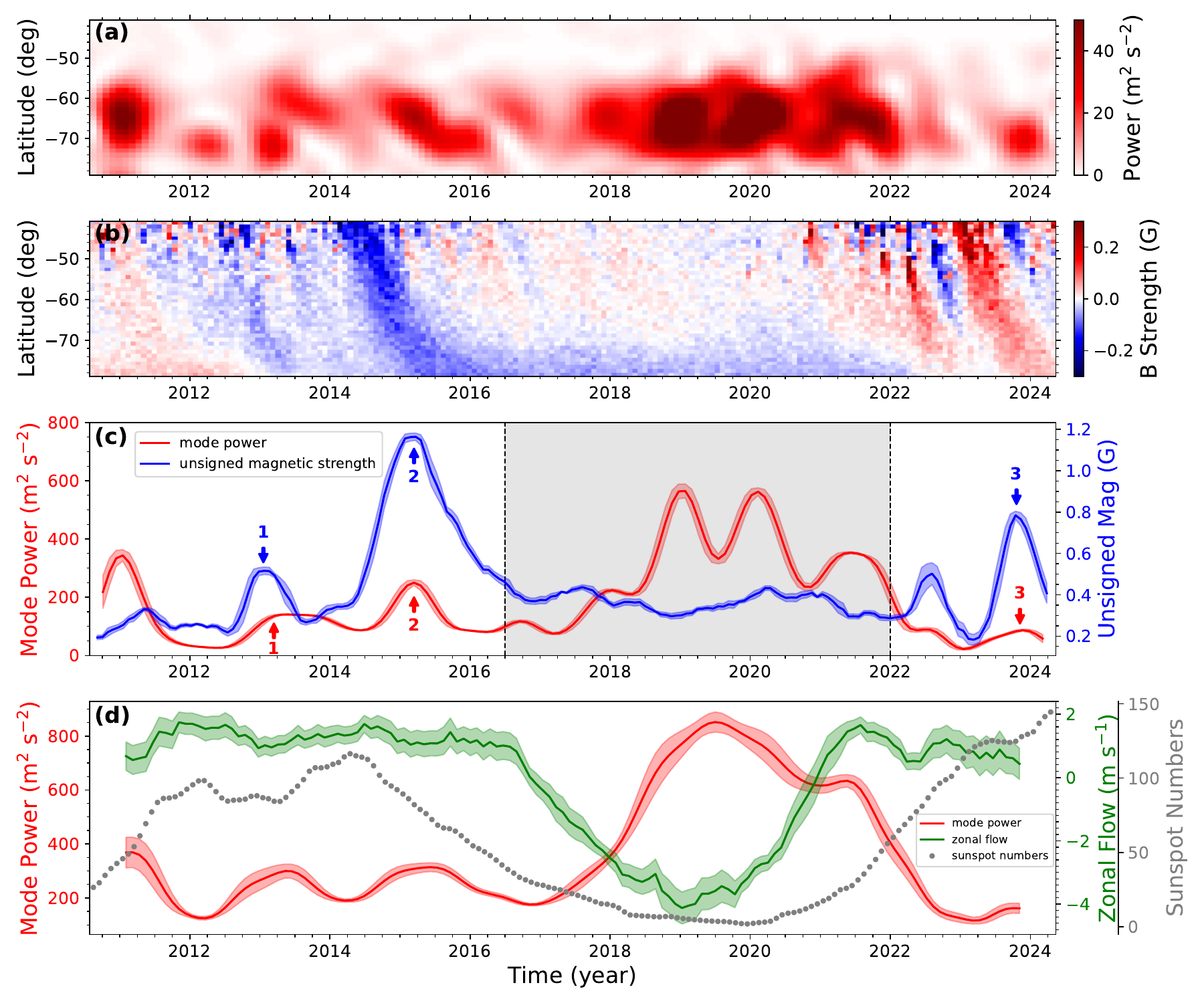}
\caption{Same as Figure~\ref{fig:n_power_t} but for the southern hemisphere.
\label{fig:s_power_t}}
\end{figure}

Using the filtered data as shown in Figure~\ref{fig:filtered_data}, we integrate the $m=1$ flow power across all longitudes for each rotation and obtain the total power of the mode for all latitudes.
Panels (a) of Figures~\ref{fig:n_power_t} and \ref{fig:s_power_t} show the temporal evolution of the $m=1$ mode power in high latitudes, revealing that in each polar region, the power fluctuates over time and tends to break into distinct mode packets.
These power breakings coincide with where phase jumps occur (see Section~\ref{sec31}), marking the end of an old mode and the start of a new one.

We compare the temporal evolution of the mode power with the evolution of the magnetic field, commonly known as the magnetic butterfly diagram, in both the northern and southern hemispheres (see Figures~\ref{fig:n_power_t}b and \ref{fig:s_power_t}b).
The temporal evolution of the mode power can be divided into two periods based on magnetic activity levels: the activity minimum period and the magnetically active period.
During the activity minimum period, approximately between 2016.5 and 2022.0, the polar magnetic field remains stable and relatively strong with one magnetic polarity. 
In this period, the mode power also remains stronger than in other time periods, although significant magnitude variations occur.
During the magnetically active periods, namely 2010.5 -- 2016.5 and 2022.0 -- 2024.2, magnetic fluxes are observed to be transported from mid-latitude to high-latitude regions. 
There is a tendency for mode breakings to occur when magnetic fluxes are transported poleward and pass the latitudes around $55\degr - 60\degr$, while the mode power strengthens when the magnetic field enhances around latitude $70\degr$.

To further quantify this relationship, panels (c) of Figures~\ref{fig:n_power_t} and \ref{fig:s_power_t} compare the mode power (integrated over $60\degr - 70\degr$) with the magnetic flux (integration of the unsigned magnetic field over $65\degr - 75\degr$), both of which are smoothed with a 5-rotation running window.
Outside the shaded area representing the activity minimum period, three local peaks of mode power and unsigned magnetic flux are marked in both hemispheres, highlighting their good correspondence in the timing of occurrence. 
This correspondence in timing, though not in magnitude, suggests that the magnetic field significantly influences the $m=1$ mode power, which gets boosted as the local magnetic field strengthens.

In addition to comparing the mode power with the high-latitude magnetic field, it is also useful to examine its relationship with the Sun’s global-scale activity. 
Panels (d) of Figures~\ref{fig:n_power_t} and \ref{fig:s_power_t} present a comparison between the total mode power (integrated over $50\degr - 80\degr$) and the monthly sunspot number\footnote{Data are downloaded from \url{https://www.sidc.be/.}}, both smoothed using a 13-month window, as done by \citet{Waidele23}. 
As shown in both polar regions, the mode power exhibits a clear anti-correlation with the sunspot number, becoming weaker during solar maximum years and stronger during minimum years, consistent with the recent finding of \citet{Liang24}. 
Notably, this trend contrasts with the behavior of the equatorial Rossby waves, which instead show a positive correlation with the sunspot number \citep{Waidele23}.

In addition to examining the relationship between the mode power and the Sun's magnetic activity, we also investigate the relationship between the mode power and the zonal flow speed.
As introduced in Section~\ref{sec2}, the mean zonal flow, corresponding to the $m=0$ component in each rotation, is removed from the synoptic flow maps before the data are filtered for the mode analysis.
As shown in panels (d) of Figures~\ref{fig:n_power_t} and \ref{fig:s_power_t}, the zonal flow velocities, averaged over $55\degr - 65\degr$ and smoothed over a 13-month period for consistency with the mode power curve, clearly slow down during activity minimum years, when the mode power substantially enhances.
There is a tendency for the mode power to be anti-correlated with the zonal flow speed. 

It is important to note that, although positive or negative correlations are observed between the aforementioned physical variables, these variables may or may not be causally related to one another, as they could all be affected or even determined by other factors. 
Further modeling efforts and investigations are needed to better understand these relationships.

\subsection{Zonal and Meridional Phase Speeds}
\label{sec33}
As shown in the animation associated with Figure~\ref{fig:filtered_data}, the $m=1$ mode remains largely stationary relative to the tracking, exhibiting only minor motions most of the time. 
The most significant motions typically occur during phase jumps when one mode transitions to another.
Characterizing these motions in both the zonal and meridional directions is valuable, as they may be associated with the local zonal and meridional flow speeds, the latter of which is notoriously challenging to measure at such high latitudes.

\begin{figure}[!t]
\plotone{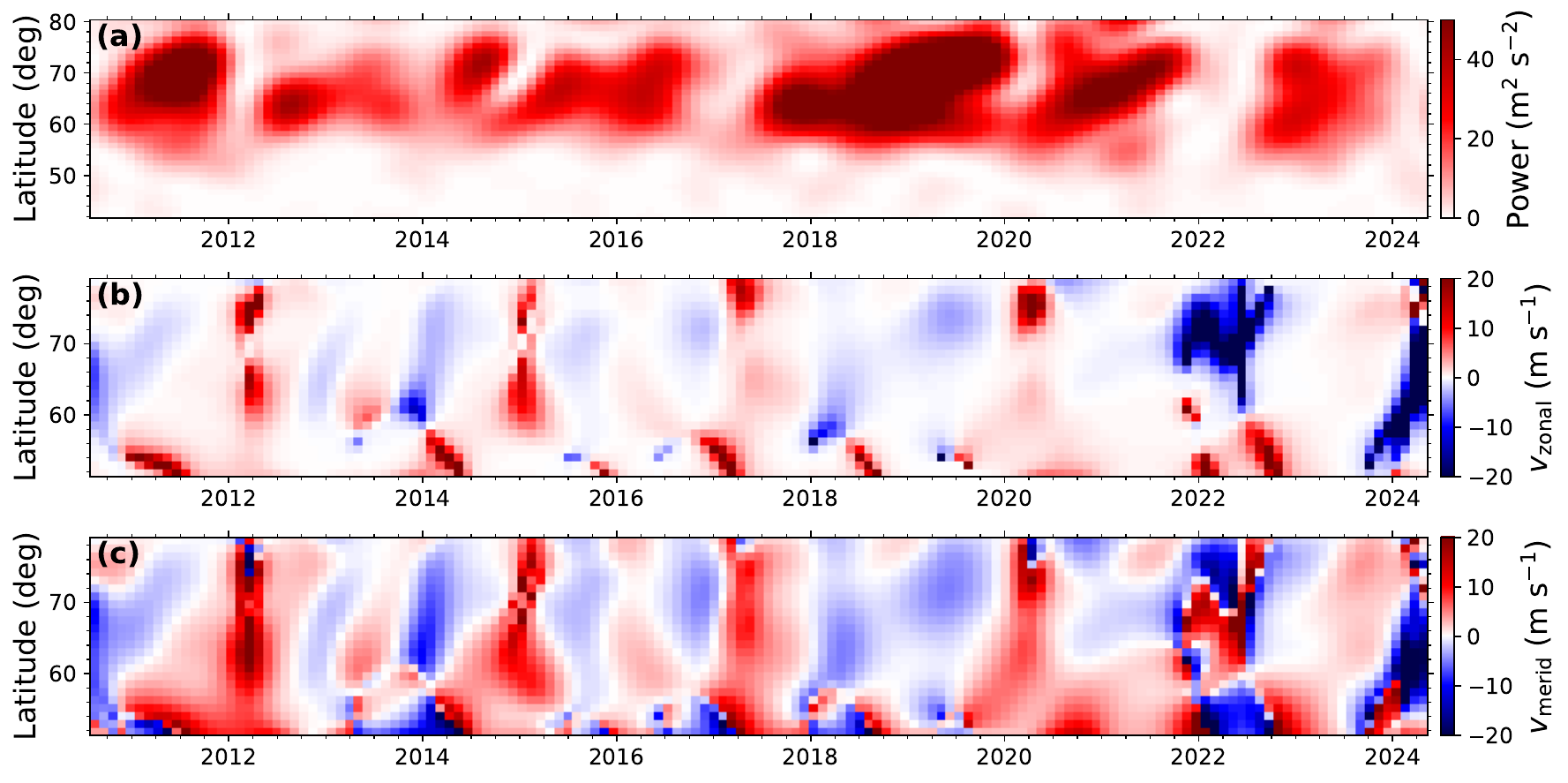}
\caption{(a) Same as Figure~\ref{fig:n_power_t}a. 
It is repeated here for a direct comparison of position and time with the zonal and meridional phase velocities in panels (b) and (c).
(b) Zonal phase velocities $v_\mathrm{zonal}$, with positive as prograde and negative as retrograde. 
(c) Meridional phase velocities $v_\mathrm{merid}$, with positive pointing north and negative pointing south.
These results are for the northern polar region.
\label{fig:n_speed}}
\end{figure}

\begin{figure}[!t]
\plotone{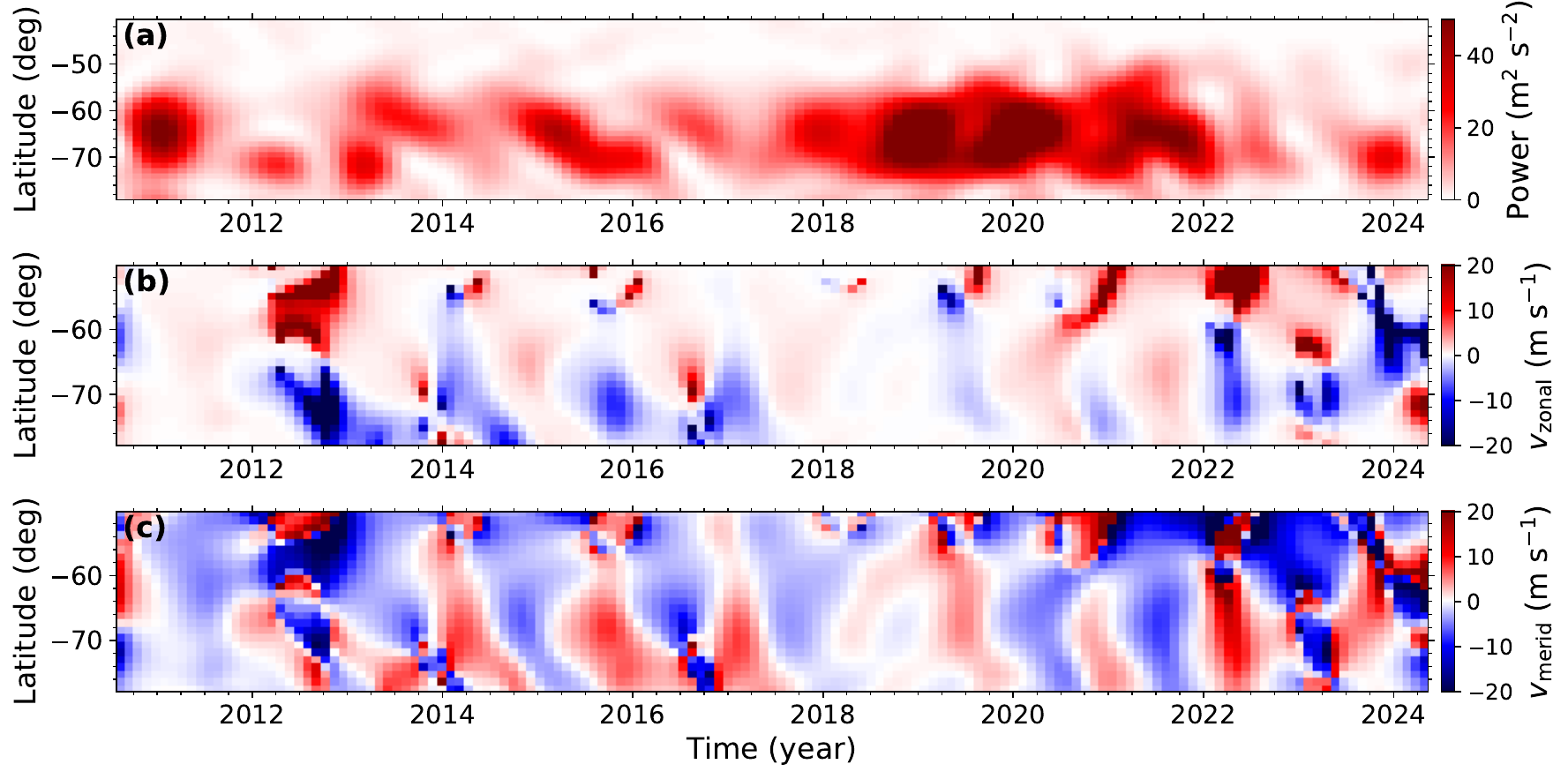}
\caption{Same as Figure~\ref{fig:n_speed} but for the southern polar region.
\label{fig:s_speed}}
\end{figure}

We would like to point out that, although the $m=1$ mode seems to be the sole mode remaining in the filtered data, it is actually composed of $\ell$'s from 0 to 20, which effectively causes the mode to behave like waves. 
Before we calculate the phase speeds of the mode, we also note that the power maps in Figures~\ref{fig:n_speed}a and \ref{fig:s_speed}a show a tendency for the power to migrate toward the poles during most of the modes' lifetimes. 
Although a precise determination is difficult using the available data, this is a strong indication that the mode's latitudinal component of group velocity is directed poleward, transporting the mode power toward the poles as the mode evolves.

We then calculate the phase velocities.
Across all longitudes at each latitude for each rotation, the filtered $m=1$ mode exhibits one period of wave pattern following the form of $\sin (\phi + \varphi_0)$.
We adopt the angle $\varphi = \phi + \varphi_0$ in the sine function as the phase of the wave, thus getting a phase array of $\varphi (\theta, \phi, t)$ for all rotations.
The phase velocity, $\vec{v}_\mathrm{ph} (t) = \omega/ k \cdot \hat{k}$, can be computed through $\omega = - \partial \varphi / \partial t$ and $\vec{k} = \nabla \varphi$ separately.
The $\hat{k}$ denotes a unit vector of the wavenumber, and $\nabla$ is calculated in spherical coordinates. 
For each rotation, $\vec{v}_\mathrm{ph} (t)$ can be decomposed into the directions parallel to and perpendicular to the solar rotation, which are zonal phase velocity $v_\mathrm{zonal} (t)$ and meridional phase velocity $v_\mathrm{merid}(t)$, respectively. 
However, one needs to be cautious that the $v_\mathrm{zonal}$ and $v_\mathrm{merid}$ may not be cleanly decoupled.
As shown in Figure~\ref{fig:filtered_data}, the $m=1$ exhibits yin-yang structures around the poles, which are displayed as inclined bands in longitude-latitude coordinate or as spiraling patterns in polar coordinate.
This morphology complicates the calculation of zonal and meridional phase velocities, as it introduces an ambiguity between the two: a prograde (retrograde) zonal motion can be interpreted as a poleward (equatorward) meridional motion, and vice versa. 
Therefore, we need to keep in mind that there is a possibility of leakage from one velocity to the other in the results presented below.

Also, please note that the mode's zonal phase velocity $v_\mathrm{zonal}$ should not be confused with the $v_\phi (\theta, \phi)$, the longitudinal (or zonal) velocities derived from the time-distance helioseismology and used to study the inertial modes. 
The $v_\phi (\theta, \phi)$ are tracked with the Sun's local rotation rate and are used to show the inertial modes patterns, and the zonal phase velocity measured here represents the mode's phase motion relative to the tracking frame. 
It is also important to note that the measured meridional phase velocity $v_\mathrm{merid}$ is expected to include both the local meridional flow and the mode's intrinsic meridional phase velocity.

Panels (b) and (c) of Figures~\ref{fig:n_speed} and \ref{fig:s_speed} illustrate the calculated $v_\mathrm{zonal} (t)$ and $v_\mathrm{merid} (t)$, respectively. 
Since $v_\mathrm{zonal}$ and $v_\mathrm{merid}$ are components of the phase velocity $\vec{v}_\mathrm{ph}$, they exhibit strong correlations in the northern polar region and anti-correlation in the southern. 
The most prominent features in both panels are the significantly larger phase speeds during the mode breakings or phase jumps, compared to those during the mode's stable lifetime. 
This is expected, as phase jumps occurring over short time intervals lead to large phase velocities. 
Within each mode’s lifetime, we highlight the following three points. 

First, for most, but not all, of the modes observed during the analysis period, the $v_\mathrm{zonal}$ tends to be prograde (positive) during the early phases of their lifetimes and becomes retrograde toward the later phases. 
Similarly, $v_\mathrm{merid}$ is generally directed poleward (positive in the northern hemisphere and negative in the southern) in the early phases, transitioning to an equatorward direction in the later phases. 
Both $v_\mathrm{zonal}$ and $v_\mathrm{merid}$ exhibit a clear latitudinal dependence, with the above-described trends appearing more pronounced at lower latitudes than at higher latitudes.

\begin{figure}[!t]
\plotone{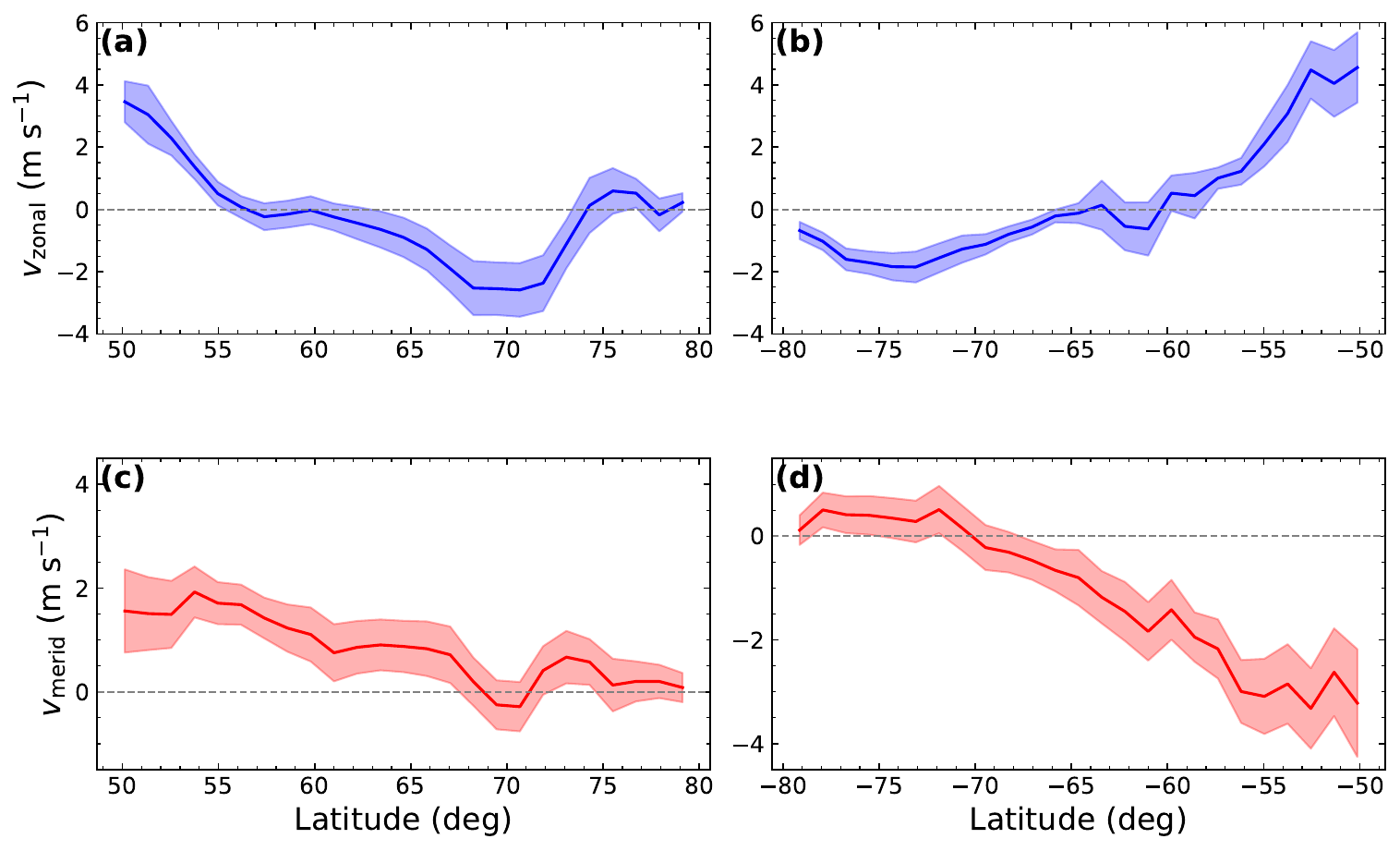}
\caption{(a) Mean zonal phase speed $\langle v_\mathrm{zonal} \rangle$ for the northern polar region, obtained by averaging the $v_\mathrm{zonal}$ of the full analysis period. Positive speeds are prograde.
(b) Same as (a) but for the southern polar region. 
(c) Mean meridional phase speed $\langle v_\mathrm{merid} \rangle$ for the northern polar region, obtained by averaging the $v_\mathrm{merid}$ of the full analysis period. Positive speeds are northward. 
(d) Same as (c) but for the southern polar region.
\label{fig:ph_speed_lat}}
\end{figure}

Second, although the entire polar region is tracked using a uniform rotational rate corresponding to latitude $65\degr$, the mode’s zonal phase velocity $v_\mathrm{zonal}$ exhibits only a weak differential rotation. 
Figure~\ref{fig:ph_speed_lat}a and ~\ref{fig:ph_speed_lat}b display the $\langle v_\mathrm{zonal} \rangle$ averaged from the whole period, which reveals a slowdown of merely $\sim6$\,m\,s$^{-1}$ across the examined latitudinal range. 
For comparison, the Sun’s rotation speed decreases by approximately 126\,m\,s$^{-1}$ over the same latitudinal range \citep{Komm1993}. 
This suggests that the $m=1$ mode rotates around the polar axis in a manner close to rigid-body rotation, experiencing only minimal influence from the strong differential rotation of the surrounding plasma.

Third, as shown in Figure~\ref{fig:ph_speed_lat}c and Figure~\ref{fig:ph_speed_lat}d, the meridional phase velocity $v_\mathrm{merid}$ is directed poleward below latitude $70\degr$, consistent with the current understanding that the Sun’s surface and near-surface meridional flow is poleward up to approximately $70\degr$ \citep[e.g.,][]{Zhao14, Komm18, Mahajan21}.   
Above latitude $70\degr$, particularly in the southern polar region, the measured $v_\mathrm{merid}$ 
shows an indication of reversing direction, becoming equatorward with speeds below 0.5\,m\,s$^{-1}$.
The existence of such a high-latitude counter-flow cell has long been debated, as its accurate detection remains challenging \citep{Haber2002, Upton2012}. 
Our measurements using the $m=1$ mode offer a new diagnostic tool, complementing the traditional approaches such as correlation or feature tracking and local helioseismology, and providing a new perspective on high-latitude meridional flow. 
However, it is reminded that ambiguity remains in distinguishing $v_\mathrm{zonal}$ and $v_\mathrm{merid}$, and that $v_\mathrm{merid}$ includes contributions from both the local meridional flow and the intrinsic motion of the inertial mode. 
Resolving the ambiguity and attributing the measurements to their causes will be the foci of future studies.

\subsection{Phase Differences between Modes of Northern and Southern Polar Regions}
\label{sec34}

The symmetrized and anti-symmetrized components of the longitudinal flows from both hemispheres were previously used in the analysis of the $m=1$ mode \citep[e.g.,][]{Gizon21, Liang24}. 
However, it is unclear whether the mode in the two polar regions are symmetric, anti-symmetric, or exhibit a certain phase difference. 
It is therefore of significant interest to examine the phase differences between the northern and southern modes and their temporal evolution throughout the solar cycle.

For each rotation, we select one row of filtered data $v_\phi^\mathrm{N}(\phi)$ at the latitude $\theta$ between $50\degr - 80\degr$ in the northern hemisphere and another row $v_\phi^\mathrm{S}(\phi)$ at the same latitude but of the southern hemisphere. 
The phase difference between these two rows is then calculated using $$
\delta\varphi^\mathrm{NS} = \arg (\widehat{v_\phi^\mathrm{N}(\phi)} \ \widehat{v_\phi^\mathrm{S}(\phi)}^\dagger),$$
where $\arg$ is to find the argument of a complex number, $\widehat{v(\phi)}$ is the Fourier transform of $v(\phi)$, and $^\dagger$ is to take the conjugate of a complex number. 
This process is repeated for every row within the $50\degr - 80\degr$ latitude range, and the same analysis is performed for each rotation. 

\begin{figure}[!t]
\plotone{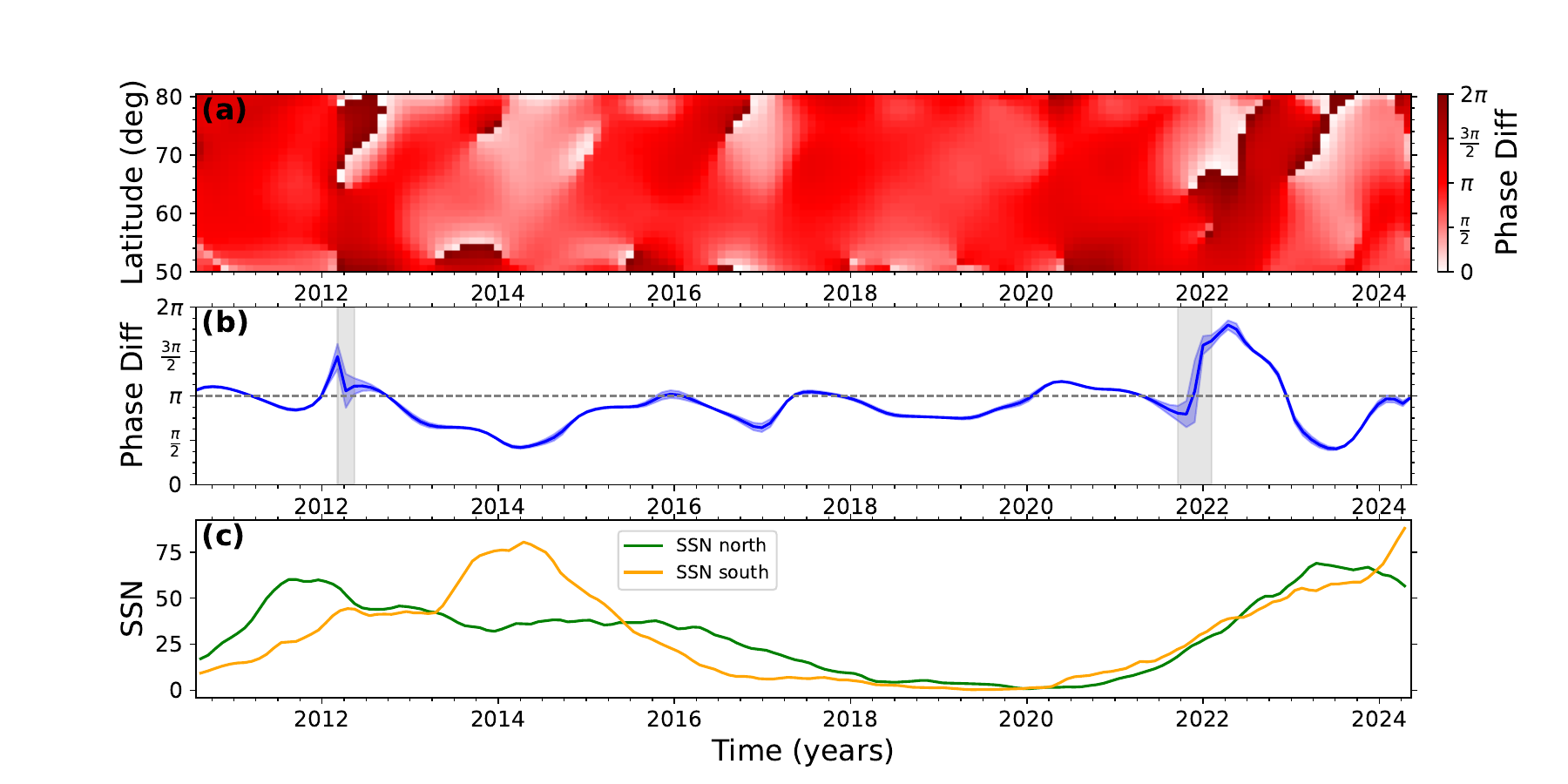}
\caption{(a) Relative phase shifts between the $m=1$ modes in the northern and southern polar regions, displayed as a function of time and latitude. 
(b) Relative phase shifts between the northern and southern polar regions, integrated over $60\degr - 65\degr$, displayed as a function of time. 
The shaded regions indicate where phase jumps occur and the measurements of phase shifts become unreliable.
(c) Mean sunspot numbers in both hemispheres, averaged over a 13-month running window.
\label{fig:n_s_phase}}
\end{figure}

Figure~\ref{fig:n_s_phase}a presents the phase difference $\delta\varphi^\mathrm{NS}$ as a function of latitude and time, where discontinuities can be seen near the time of phase jumps.
Figure~\ref{fig:n_s_phase}b shows the $\delta\varphi^\mathrm{NS}$ averaged over $60\degr - 65\degr$, the latitudinal range that is less affected by the phase jumps.
It can be seen that the $\delta\varphi^\mathrm{NS}$ varies rather smoothly, remaining close to $\pi$ throughout the analysis period except during the period of 2013 -- 2015 and briefly in 2023 when $\delta\varphi^\mathrm{NS}$ drops to approximately $\pi/2$. 
This suggests that the $m=1$ mode in each hemisphere is close to being antisymmetric most of the time. 
Even when they deviate from this state of near-antisymmetry, they gradually migrate back to it.

It is worth noting that during the period of 2013 -- 2015 when $\delta\varphi^\mathrm{NS}$ deviates substantially from $\pi$, the sunspot number in the southern hemisphere\footnote{Data are downloaded from \url{https://www.sidc.be/SILSO/monthlyhemisphericplot}} substantially exceeds that of the northern hemisphere (Figure~\ref{fig:n_s_phase}c). 
However, for most of the period when the northern hemisphere exhibits a higher sunspot number than the southern hemisphere, the phase difference between the modes in the two polar regions does not rise significantly above $\pi$. 
While this temporal coincidence is intriguing, it remains unclear whether the dominance of southern hemisphere activity plays a causal role in the observed deviation from the mode's antisymmetry or whether the correlation is merely coincidental.

\section{Discussion and Conclusion}
\label{sec4}

Using time-distance subsurface flow maps, we have analyzed the $m=1$ inertial mode near both solar poles. 
With synoptic flow maps constructed through tracking the rotation rate at latitude $65\degr$, we decompose the spherical modes and calculate the power distributions for $m=1$ and $m=2$ modes in both polar regions, revealing strong power in $m=1$ near zero frequency.
The $m=1$ mode in both polar regions exhibit characteristic yin-yang structures, with power amplitudes showing intriguing connections to both local- and global-scale magnetic activity. 
The mode power amplitude is found anti-correlated with the monthly sunspot number and mean zonal flow, remaining consistently strong during the activity minimum years when the polar magnetic field is built up and stable. 
During the magnetically active years, the mode appears to be strengthened with the arrival of magnetic flux transported from the mid-latitude regions. 
Mode breakings occur between these strengthened periods when magnetic fluxes are being transported toward where the mode resides.
We have also calculated the zonal and meridional phase speeds of the $m=1$ mode throughout its lifetime, characterizing its dynamics. 
The mode's zonal phase speed is less differential with latitude compared to its surrounding solar plasma, and the meridional phase speed, composed of local meridional flow and the mode's characteristic dynamics, is directed poleward below latitude $70\degr$ and equatorward above it. 
Interestingly, the $m=1$ mode in two polar regions remains close to antisymmetry for most of the time, but deviates from it for some short periods.

It is of particular interest to understand how the inertial modes and magnetic fields influence each other. 
Recent numerical simulations \citep{Dikpati24} suggest that magnetic flux migrating toward the poles could contribute to the formation of polar vortices, which are linked to the faster and slower longitudinal flows identified as low-$m$ inertial modes in our study. 
Indeed, our analysis reveals a strong correlation between magnetic surges in high-latitude regions and the enhancement of the mode power during the active phases of the solar cycle.
At the same time, our study also shows that during activity minimum years when the polar magnetic field is strong and stable, the inertial mode is also strong and stable with longer lifetimes than the periods with active poleward flux transport. 
Overall, the local magnetic field strength has a strong impact on the power amplitude of the $m=1$ mode and its lifetimes. 

Additionally, our analysis reveals an anti-correlation between the mode power and zonal flow speed, with stronger power observed during the periods of slower rotation. 
This suggests that slower rotation may facilitate the formation and strengthening of low-$m$ modes.
However, the opposite may also be true, as recent numerical modeling suggests, low-$m$ modes could play a key role in regulating the Sun’s differential rotation \citep{Bekki24}.
Here, we emphasize that the positive and negative correlations found in this study between different physical quantities do not necessarily imply a causal relationship. 
It is possible that all these observed variations stem from a deeper, underlying mechanism that has yet to be discovered.

The result that the modes in both polar regions exhibit phases close to antisymmetry for most of the analysis period is striking. 
Even more intriguing is that, despite brief deviations from this antisymmetry -- when the southern hemisphere experiences stronger magnetic activity -- the mode gradually migrates back to its near-antisymmetry state.
This raises important questions: What controls the phases of the $m=1$ mode in both polar regions? 
What mechanisms maintain them in an anti-phase state? 
How does the magnetic activity asymmetry between the two hemispheres influence the mode phases near the poles? 
Addressing these questions requires a deeper understanding of how these low-$m$ modes are excited in the solar interior, how they rise to the surface in the polar regions, what sustains them over their approximate two-year lifetimes, and how these modes interact with the local and global magnetic fields.

Throughout their lifetimes, the $m=1$ mode remains largely stationary relative to its surrounding plasma, yet they do not remain completely still. 
Our calculations show that at the beginning of their lifetimes, these modes tend to rotate faster, moving prograde relative to the reference frame in most cases, before they gradually slow down and become retrograde near the end of their lifetimes. 
Typically, a substantial slowdown signals the end of a mode and the emergence of a new one.
Measuring meridional flow at latitudes above $70\degr$ is notoriously challenging using either tracking or helioseismic method, and the possible existence of a counter-flow cell in the polar regions has long been a topic of scientific interest, with important implications for surface flux transport dynamo models \citep{Upton2014}.
While our measured meridional phase velocity aligns well with the expected poleward flow below $70\degr$, it exhibits an intriguing sign reversal above this latitude. 
Whether this measured equatorward velocity is due to the actual equatorward plasma flow or is due to the poorly-understood dynamics of the inertial mode remains unclear, but this question undoubtedly merits further investigation.

If the meridional flow above $70\degr$ indeed represents the local plasma motion, angular momentum would be transported away from the poles, further decelerating them.
However, as is widely known, the meridional flow is not the only mechanism responsible for transporting angular momentum; Reynolds stress, defined as the average product of the fluctuating velocity components of the latitudinal and longitudinal flows, is another contributor.
It would be useful if we can calculate the Reynolds stress near the polar region using the time-distance helioseismic subsurface flows; however, as pointed out in Section~\ref{sec2}, the latitudinal flows from this analysis pipeline are affected by the strong center-to-limb effect, making it unreliable to calculate Reynolds stress at latitudes above approximately $65\degr$.
Hopefully, a robust method can be developed to eliminate the center-to-limb effect from subsurface flows, allowing for the calculation of Reynolds stress in the high-latitude region and thereby improving our understanding of angular momentum transport in that area.

In summary, we have characterized the high-latitude $m=1$ mode in terms of its power distribution across frequency, its zonal and meridional phase speeds, the phase differences between the northern and southern polar modes, and the mode's relationships with both local and global magnetic fields, as well as with the local zonal speed.
These findings provide valuable insights into the inertial mode while also demanding a deeper understanding of their interior dynamics, which requires further theoretical and numerical modeling.

\begin{acknowledgments}
\sdo\ is a NASA mission, and HMI is an instrument developed by Stanford University under the NASA contract number NAS5-02139. 
The monthly sunspot number is downloaded from WDC-SILSO, Royal Observatory of Belgium, Brussels, DOI: \url{https://doi.org/10.24414/qnza-ac80}.
This work is partially sponsored by NASA DRIVE Science Center COFFIES project under grant number 80NSSC22M0162.
\end{acknowledgments}

\bibliography{ms}{}
\bibliographystyle{aasjournal}

\end{document}